\documentclass[aps,prb,twocolumn,superscriptaddress,showpacs]{revtex4-1}

\usepackage{amsmath}
\usepackage{amssymb}
\usepackage{graphicx}
\usepackage{multirow}
\usepackage{amsmath}
\usepackage{amssymb}

\newcommand{\ii}{\text{i}}
\newcommand{\dd}{\text{d}}
\newcommand{\ee}{\text{e}}
\newcommand{\ep}{\varepsilon}
\newcommand{\mb}{\mathbf}
\newcommand{\sgn}{\text{sign}}

\begin{document}


\title{Relaxation of optically excited carriers in graphene:\\ Anomalous diffusion and L\'evy flights}


\author{U. Briskot}
\affiliation{Institute of Nanotechnology, Karlsruhe Institute of Technology, 76021 Karlsruhe, Germany}
\affiliation{Institute for Theoretical Condensed Matter Physics and Center for Functional Nanostructures, Karlsruhe Institute of Technology, 76128 Karlsruhe, Germany}
\author{I. A. Dmitriev}
\affiliation{Max Planck Institute for Solid State Research, Heisenbergstr. 1, 70569 Stuttgart, Germany}
\affiliation{Institute of Nanotechnology, Karlsruhe Institute of Technology, 76021 Karlsruhe, Germany}
\affiliation{Institute for Theoretical Condensed Matter Physics and Center for Functional Nanostructures, Karlsruhe Institute of Technology, 76128 Karlsruhe, Germany}
\affiliation{Ioffe Physical Technical Institute, 194021 St. Petersburg, Russia}
\author{A. D. Mirlin}
\affiliation{Institute of Nanotechnology, Karlsruhe Institute of Technology, 76021 Karlsruhe, Germany}
\affiliation{Institute for Theoretical Condensed Matter Physics and Center for Functional Nanostructures, Karlsruhe Institute of Technology, 76128 Karlsruhe, Germany}
\affiliation{Petersburg Nuclear Physics Institute, 188300 St. Petersburg, Russia}

\date{\today}

\begin{abstract}
We present a theoretical analysis of the relaxation cascade of a
photoexcited electron in graphene in the presence of RPA screened electron-electron interaction.
We calculate the relaxation rate of high energy electrons
and the jump-size distribution of the random walk constituting the cascade
which exhibits fat tails.
We find that the statistics of the entire cascade are described by L\'evy flights with
constant drift instead of standard drift-diffusion in energy space.
The L\'evy flight manifests nontrivial scaling relations of the fluctuations
in the cascade time, which is related to the problem of the first passage time
of L\'evy processes.
Furthermore we determine the transient differential transmission of graphene
after an excitation by a laser pulse taking into account the fractional kinetics
of the relaxation dynamics.
\end{abstract}


\maketitle

\section{Introduction \label{sec:introduction}}
\noindent
The fabrication of graphene\cite{Novoselov2004} launched
a new era of two-dimensional (2D) materials in condensed matter physics,
giving access to fundamentally different phenomena
and systems realized for the first time in a solid state environment.\cite{Zhang2005,NovoselovQH2006,Bolotin2008,Du2008}
Graphene promises to be an attractive platform for electronic\cite{Castro2009}
and in particular optoelectronic applications,\cite{Bonaccorso2010,Avouris2010,Engel2012}
where research reaches from lasing\cite{PopInv2012} to energy conversion.\cite{Gabor2011,SongHCTrans2011}
The nature of interactions and their interplay will limit the intrinsic properties
of graphene devices and has therefore attracted interest from the application-oriented as well
as fundamental standpoint.
For the latter, neutral or intrinsic graphene embodies the paradigm of a
marginal Fermi liquid (FL).\cite{SheehySchmalian2007,Gonzales1996,FritzQCTrans2008} While graphene in the presence of
electron-electron interactions (EEI) establishes a finite Fermi surface at high doping,
it crosses over to a relativistic Dirac liquid at lower densities and manifests
non-FL relaxation rates\cite{DasSarmaFLnFL2007,DasSarmaScreening2007,Schuett2011}
and transport characteristics.\cite{FritzQCTrans2008,Castro2009,Kashuba2008}
Another interesting interaction-dominated transport phenomenon is Coulomb drag in graphene double layer systems
\cite{KimDrag2011,GorbachevDrag2012,GeimSchuettDrag2013} which is determined by the peculiar interaction-induced interlayer relaxation.
\cite{DasSarmaDrag2007,KatsnelsonDrag2011,BorisDrag2012,FritzDrag2012,LevitovDrag2013,SchuettDrag2013}
In the last years it became feasible to examine the interactions even on very short time scales
by means of ultra-fast pump-probe measurements.\cite{Sun2008,Dawlaty2008,Winnerl2013}
They revealed that EEI in graphene dominates over phonon interaction at an early stage of
relaxation processes making graphene a highly efficient material for thermoelectric applications.\cite{Breusing2009,Shang2010}
On the other hand the relaxation of high energy electrons follows again a non-FL scheme
as electrons relax via a cascade of small steps in energy space.\cite{TielrooijKoppens2013}

So far theoretical work focused on the relaxation rates of thermal electrons
using static screening or dynamical screening
in the random phase approximation (RPA).
\cite{DasSarmaScreening2007,Schuett2011,Gonzales1996,DasSarmaFLnFL2007,Kashuba2008,FritzQCTrans2008,HwangDasSarma2007,Polini2007,Polini2008,Ramezanali2009}
Comprehensive numerical studies elucidated the interplay of EEI and
phonon interactions\cite{BreusingMalic2011,Avouris2011} as well as the
importance of different scattering channels in particular in the context of
carrier multiplication via Auger processes.\cite{Tomadin2013}
The influence of flexural phonons\cite{vonOppenFlexPhonons08,*vOppenFlexPhonon2009,GornyiFlexPhonons2012}
in free-standing graphene
and combined effects of phonons and disorder\cite{SongSuperColl2012} have been studied in detail.
The relaxation of optically excited carriers in doped graphene\cite{Brida2013,TielrooijKoppens2013}
was theoretically studied\cite{SongKoppens2013} at zero temperature and is consistent with the cascade picture.

In this work we present an analysis of the relaxation cascade at finite temperature.
We consider the first stage of the relaxation process dominated by electron-electron
collisions and neglect phonon and disorder effects.
In Sec.~\ref{sec:cascade_step} we study a single cascade step for undoped as well as for doped graphene
in Sec.~\ref{sec:cascade_step} and calculate the relaxation rates of high energy
electrons in graphene using RPA. The main result of Sec.~\ref{sec:cascade_step} is the
distribution of the size of a single jump in the random walk describing the relaxation cascade.
In Sec.~\ref{sec:cascade} we infer the characteristics of the whole cascade
on the basis of the results presented in Sec.~\ref{sec:cascade_step}, with emphasis
on the fluctuations on top of the particle's drift in energy space.
The cascade process manifest the unique Dirac nature of carriers in graphene
as it is described by L\'evy flights.\cite{Feller1971}
Finally, in Sec.~\ref{sec:FFPE} we determine the transient differential transmission
of a graphene sample after excitation with a laser pulse in the presence of EEI.

\section{Single cascade step \label{sec:cascade_step}}
\noindent
We are going to discuss the relaxation of carriers excited by a laser pulse with
central frequency $\omega_\text{pump}$.
We focus on the dynamics of the excited electrons rather than the questions
associated with the equilibration of the low energy thermal electrons.
We restrict our analysis to the earliest stage dominated by EEI, 
in which the energy remains entirely in the electronic system.
For moderate pump fluence the phase space density of the excited electrons is
much lower then the one of thermal electrons. Scattering and energy relaxation of
a high energy excited electron is therefore predominantly due to interaction
with thermal electrons. We neglect the mutual scattering of high energy electrons
and assume that the low energy electrons remain thermal with temperature $T$.
For small fluences we also neglect the change in $T$ due to illumination.
In this sense the excited electrons with an energy of the order
$\omega_\text{pump}/2$ are relaxing in consecutive steps due to the interaction with a
thermal bath of low energy electrons at equilibrium.

\begin{figure}[h]
	\includegraphics[width=\linewidth]{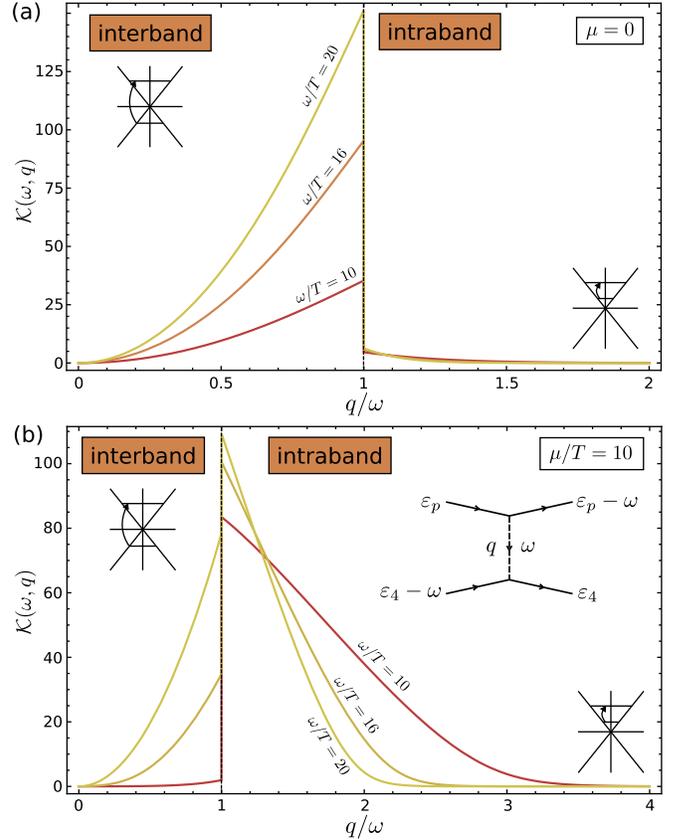}
	\caption{(Color online) The kernel $\mathcal{K}(\omega,q)$,
	Eqs.~\eqref{sec:cascade_step:eq:K_definition} and \eqref{app:kinKernel:eq:kin_Kernel},
	determining the phase space of scattering for thermal electrons
	for different frequencies $\omega$ and (a) $\mu=0$, (b) $\mu/T=10$.
	The regions of intraband ($q>|\omega|$) and interband ($q<|\omega|$) scattering are separated by the dashed line.
	\label{sec:cascade_step:fig:K}}
\end{figure}

In the following we label the eigenstates $|\lambda,\vec{k}\rangle$ of the
graphene Hamiltonian $H_0=v_F\vec{\sigma}\cdot\vec{k}$ with energy
$\ep_{\lambda k}=\lambda v_F k$ by the momentum $\vec{k}$ and band index
$\lambda=\pm 1$.
In the following we set $v_F=\hbar=1$.
We define the relaxation rate via the semiclassical Boltzmann equation
\begin{equation}
	\partial_t f_\lambda(\vec{k}) = \textit{St}[f_\lambda(\vec{k})] \: .
	\label{sec:cascade_step:eq:BE}
\end{equation}
Here $f_\lambda(\vec{k})$ is the occupation of the state $|\lambda,\vec{k}\rangle$.
The collision integral $\textit{St}[f]$ describes the electron-electron scattering.
Based on the approximations mentioned above we follow the evolution of a single
excited electron starting at momentum $\vec{p}$ as it relaxes due to scattering
with the thermal electrons with energies $\ep\ll \ep_p$. We make the ansatz
\begin{equation}
	f_\lambda(\vec{k}) = f_T(\lambda k) + \delta f_\lambda(\vec{k}) \: , \quad
	\delta f_\lambda(\vec{k}) = \delta_{\lambda,+1}\delta_{\vec{k},\vec{p}} \: .
	\label{sec:cascade_step:eq:ansatz_f}
\end{equation}
Here $f_T(\ep)=1/[1+\exp((\ep-\mu)/T)]$ is the Fermi-Dirac distribution.
With the ansatz~\eqref{sec:cascade_step:eq:ansatz_f}, the relaxation rate
of the high energy electron is determined by the outscattering rate
in the collision integral,
\begin{equation}
	\textit{St}[f_{+1}(\vec{p})] = 
	- \sum_{2,3,4} \:
	W_{12,34} f_3(1-f_4) (1-f_2) \: .
	\label{sec:cascade_step:eq:out_scattering}
\end{equation}
In Eq.~\eqref{sec:cascade_step:eq:out_scattering} we used the short-hand notation
$i=(\ep_i,\vec{k}_i)$. The transition rate $W_{12,34}$ is given in
App.~\ref{sec:app_BE}. Here, we only want to point out that in the case of Dirac
particles it contains the overlap of the eigenstates
$\langle \lambda_j, \vec{k}_j | \lambda_i, \vec{k}_i \rangle$,
that leads to a suppression of backscattering,
in addition to the semiclassical matrix element of Coulomb scattering.
In terms of the transfered energy $\omega$ and momentum $q$,
$\ep_2=\ep_p-\omega$, $\ep_3=\ep_4-\omega$ and $\vec{k}_2=\vec{p}-\vec{q}$,
$\vec{k}_3=\vec{k}_4-\vec{q}$, due to the conservation of energy and momentum,
see inset in Fig.~\ref{sec:cascade_step:fig:K}(b).

We can classify the possible scattering processes in terms of interband, $|\omega|>q$
and intraband scattering, $|\omega|<q$.
Collinear scattering occurs exactly at $|\omega|=q$.

Combining Eqs.~\eqref{sec:cascade_step:eq:BE} and \eqref{sec:cascade_step:eq:out_scattering}
we obtain an expression for the relaxation rate $\Gamma(p)$ of the photoexcited electron,
defined by the Boltzmann equation
\begin{equation}
	\partial_t f_{+1}(\vec{p}) = - \Gamma(p) = \textit{St}[f_{+1}(\vec{p})] \: ,
	\label{sec:cascade_step:eq:BE_Gamma}
\end{equation}
which is written as
\begin{equation}
	\Gamma = \int_{-\infty}^{+\infty}\dd\omega \: P(\omega) \: .
	\label{sec:cascade_step:eq:Gamma_P}
\end{equation}
Here $P(\omega)$ is the scattering rate per frequency interval $(\omega,\omega+\dd\omega)$.
On the other hand it defines the distribution of the transfered energy
in a single scattering event or cascade step. We thus refer to $P(\omega)$ as the
jump-size distribution (JSD) of the relaxation cascade.

As long as $\omega<\ep_p$ the excited electron is scattered within the conduction
band, which implies $q>|\omega|$. Since the particle number in the conduction
and valence band are separately conserved in pair collisions,
the thermal electron that scatters with the high energy electron
also performs an intraband transition.\cite{FosterAleiner2009}
We find that the contribution for $\omega>\ep_p$ corresponding to interband
transitions is negligible for the relaxation rate $\Gamma$,
Eqs.~\eqref{sec:cascade_step:eq:BE_Gamma} and \eqref{sec:cascade_step:eq:Gamma_P},
as well as for the statistics of the entire cascade (see Sec.~\ref{sec:cascade}).
Moreover, calculation shows that the relevant transfered energies satisfy $|\omega|\ll\ep_p$.
Scattering in this case is predominantly in forward direction,
which simplifies the overlap functions
\begin{equation}
	|\langle \lambda_2, \vec{k}_2 | +1, \vec{p} \rangle|^2
	= \frac{1+\lambda_2(\vec{p}\cdot\vec{k}_2)/p k_2}{2} \simeq 1 \: .
	\label{sec:cascade_step:eq:Dirac_factor}
\end{equation}
Taking into account that $f_2\simeq 0$ for $|\ep_p-\omega|\gg\max(|\mu|,T)$ in
Eq.~\eqref{sec:cascade_step:eq:out_scattering},
we obtain the compact expression for the JSD,
\begin{equation}
	P(\omega) = \int_{|\omega|}^{\infty} \dd q \:
	q \: \frac{N|V(\omega,q)|^2}{|q^2-\omega^2|} \: \mathcal{K}(\omega,q) \: .
	\label{sec:cascade_step:eq:P_definition}
\end{equation}
Here we assumed $\ep_p\gg\max(|\mu|,T)$ and as a consequence $P(\omega)$ is independent
of the particle energy $\ep_p$.
In Eq.~\eqref{sec:cascade_step:eq:P_definition} the RPA-screened matrix element of Coulomb scattering
\begin{equation}
	V(\omega,q) = V_0(q)/\ep(\omega,q) \: ,
	\label{sec:cascade_step:eq:coul_kernel}
\end{equation}
where the dielectric function $\ep(\omega,q)=1+V_0(q) N \Pi(\omega,q)$.
The RPA polarization operator $\Pi(\omega,q)$ is given in App.~\ref{sec:app_RPA}
and the bare Coulomb interaction $V_0(q)=2\pi\alpha_g/q$.
The number of flavors $N=4$ and the
coupling constant in graphene $\alpha_g=e^2/\epsilon \hbar v_F$
in our notations is $\alpha_g=e^2/\epsilon$.
Note that in the presence of a dielectric environment with dielectric constant $\epsilon\gg1$
the coupling constant can be small, $\alpha_g\ll 1$, which we assume in the following.
The kernel
\begin{equation}
	\begin{split}
		\mathcal{K}(\omega,q) = & \int_{-\infty}^{+\infty}\dd\ep_4 \:
		\sqrt{(\omega-2\ep_4)^2-q^2} \\
		& \times f_T(\ep_4-\omega)[1-f_T(\ep_4)] \: ,
	\end{split}
	\label{sec:cascade_step:eq:K_definition}
\end{equation}
expresses the phase space (for $q>|\omega|$) of the thermal electrons
that scatter with the high energy photoexcited electron.

Let us briefly comment on the validity of the RPA. For small frequencies, the RPA
sums up the leading logarithmically divergent diagrams.\cite{Schuett2011}
For $|\omega|>\max(T,|\mu|)$, however, the RPA is justified by a large $N$ expansion.
By the same degree of approximation we also neglected the exchange term
in the collision integral.

We observe that the denominator of the integrand in Eq.~\eqref{sec:cascade_step:eq:P_definition}
is singular in the case of collinear scattering $|\omega|=q$,
which in the absence of screening would lead to the logarithmically divergent Coulomb
scattering integral.\cite{Landau10,Sachdev1998,FritzQCTrans2008}
However the polarization operator in RPA is also divergent in the case of collinear scattering,
thus the total scattering amplitude remains finite.
The singular nature of the scattering of Dirac particles with linear dispersion also
manifest itself in the phase space kernel~\eqref{sec:cascade_step:eq:K_definition}.
Figure~\ref{sec:cascade_step:fig:K} shows $\mathcal{K}$ for intrinsic graphene
($|\mu|\ll T$) as well as for $|\mu|\gg T$. In either case $\mathcal{K}$ exhibits a jump
at collinear scattering. One observes that for $\mu=0$ [Fig.~\ref{sec:cascade_step:fig:K}(a)]
the phase space of intraband processes is strongly suppressed and controlled by $T$.
On the contrary, for $|\mu|\gg T$ [Fig.~\ref{sec:cascade_step:fig:K}(b)]
$\mathcal{K}$ is dominated by intraband processes.

Below we discuss the JSD separately for $T\gg|\mu|$ and $|\mu|\gg T$.

\subsection{The limit $T\gg|\mu|$}
\begin{figure}[h]
	\includegraphics[width=\linewidth]{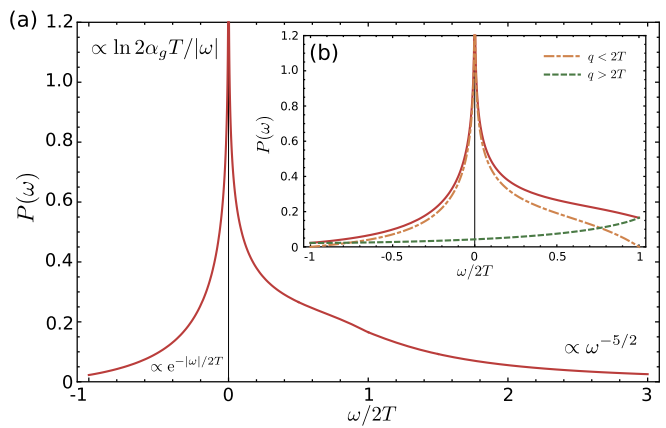}
	\caption{(Color online) The jump-size distribution~\eqref{sec:cascade_step:eq:P_definition}
	for $T\gg|\mu|$. The inset (b) shows the contributions of $q>2T$ (dashed line)
	and $q<2T$ (dash-dotted line) to $P(\omega)$ (solid line) for $|\omega|<2T$.
	Both curves are calculated for $\alpha_g=0.75$.
	\label{sec:cascade_step:fig:P_mu_zero}}
\end{figure}

\noindent
For $T\gg|\mu|$, there are two important scattering processes.
The first one is intraband scattering with small momentum transfer $q<2T$,
which leads to a logarithmic divergence in the JSD
for frequencies $|\omega|<\alpha_g T$, depicted as the dash-dotted line
in Fig.~\ref{sec:cascade_step:fig:P_mu_zero}(b).
The logarithm occurs due to the failure of screening at small frequencies and momenta
which enables resonant forward scattering.
It is the only surviving feature of the logarithmic divergence of
the unscreened Coulomb scattering integral typical for 2D systems.
The contribution of scattering with $q<2T$ decreases monotonically
with increasing frequency and vanishes for $|\omega|\ge 2T$
since $|\omega|>q$ forbids intraband scattering.

The second kind of process is intraband scattering with large momentum transfer $q>2T$.
This contribution increases with increasing frequency up to $\omega=2T$.
It dominates over scattering with small momentum transfer for $\omega\sim 2T$
and higher frequencies.
For frequencies $\omega>2T$ it decreases monotonically. Specifically,
we find that at large $\omega$ the JSD falls of as $\omega^{-5/2}$,
shown in Fig.~\ref{sec:cascade_step:fig:P_mu_zero}(a).
There is a finite probability for the excited electron to gain energy from the
bath of thermal electrons. However negative frequencies are exponentially
suppressed as shown in Fig.~\ref{sec:cascade_step:fig:P_mu_zero}(a).
The slow decay of the JSD for large frequencies has important implications for the fluctuations
of $\omega$ as discussed in Sec.~\ref{sec:cascade}. In particular it is different
from the JSD of a FL which is flat in the range $0<\omega<\ep_p$.
Thus an electron in a FL would lose most of its energy by a single jump.
The FL regime is realized under the conditions $|\mu|\gg T$ and $\ep_p\ll|\mu|$.

It turns out that for the scattering rate~\eqref{sec:cascade_step:eq:Gamma_P}
the region $|\omega|<2T$ is most important and
\begin{equation}
	\Gamma = \kappa \alpha_g T \,,\qquad |\mu|\ll T,\;\alpha_g\ll1 \: ,
	\label{sec:cascade_step:eq:Gamma_mu_zero}
\end{equation}
where $\kappa=4 \pi^2 (1+\ln 2 + G/2)\simeq 84.92$ and $G\simeq0.916$ is the Catalan constant.
The linear dependence on $T$ is a characteristic feature of intrinsic graphene
that distinguishes it from the FL.\cite{Gonzales1996}
Furthermore, due to screening the rate~\eqref{sec:cascade_step:eq:Gamma_mu_zero}
is independent of the number of flavors $N$ and linear in $\alpha_g$ contrary
to the golden rule result $\Gamma\propto\alpha_g^2 T$.\cite{Schuett2011}
The rate~\eqref{sec:cascade_step:eq:Gamma_mu_zero} is also independent of
the particle energy $\ep_p\gg\max(|\mu|,T)$.

\subsection{The limit $|\mu|\gg T$}
\begin{figure}[h]
	\includegraphics[width=\linewidth]{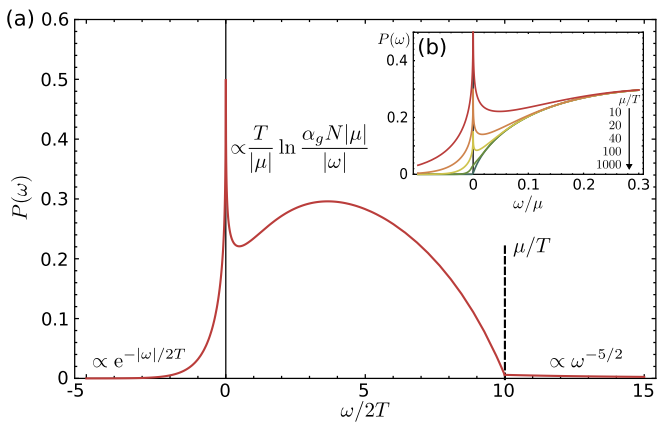}
	\caption{(Color online) The JSD~\eqref{sec:cascade_step:eq:P_definition} for $\mu/T=10$.
	In the region $|\omega|<2|\mu|$ processes with $q<2|\mu|$ are dominant. For $|\omega|>2|\mu|$
	processes with $q>2|\mu|$ determine the fat tail of the JSD.
	The inset (b) illustrates the evolution of the
	forward scattering resonance with lowering temperature.
	Both curves are calculated for $\alpha_g=0.75$.
	For details of the calculation, see App.~\ref{sec:app_BE}.
	\label{sec:cascade_step:fig:P_mu_finite}}
\end{figure}

\noindent
For $|\mu|\gg T$ the JSD is dominated by the region $|\omega|<2|\mu|$
as can be seen in Fig.~\ref{sec:cascade_step:fig:P_mu_finite}(a) while
the weight of the tail is strongly reduced.
In particular the mean jump-size will be of the order $|\mu|$.
At the lowest frequencies $|\omega|<\alpha_g T$, the JSD $P(\omega)$
shows a logarithmic divergence due to unscreened collinear scattering.
Here the JSD recovers the FL form $P(\omega)\propto (T/|\mu|)\ln|\mu/\omega|$
[see Ref.~\onlinecite{DasSarmaFLnFL2007}]
in contrast to the result for $T\gg|\mu|$, where we obtain $P(\omega)\propto\ln(\alpha_g T/|\omega|)$.
In the $T=0$ limit the logarithmic divergence at small energies
vanishes, see Fig.~\ref{sec:cascade_step:fig:P_mu_finite}(b).
In this case $P(\omega)$ reproduces the result of Ref.~\onlinecite{SongKoppens2013}.

The dominant process for $|\omega|<2|\mu|$ is the intraband scattering
with small momentum transfer, $q<2|\mu|$.
Similar to the case $T\gg|\mu|$, such small-momentum scattering
is not possible for $\omega>2|\mu|$ where
scattering with $q>2|\mu|$ leads to the fat tail $\propto\omega^{-5/2}$.
The contribution of negative frequencies $P(\omega<0)\propto\exp(\omega/2T)$ is exponentially small.

In the case $T\gg|\mu|$, the relaxation rate was determined by $|\omega|<2T$.
The total rate for $|\mu|\gg T$, is dominated by $0<\omega<2|\mu|$ and is given by
\begin{equation}
	\Gamma=8\alpha_g\pi^2 |\mu|\,,\qquad |\mu|\gg T,\;\alpha_g\ll1 \: .
	\label{sec:cascade_step:eq:Gamma_mu_finite}
\end{equation}

The rates~\eqref{sec:cascade_step:eq:Gamma_mu_finite} and \eqref{sec:cascade_step:eq:Gamma_mu_zero}
are calculated in the ballistic regime $T\tau_\text{dis}\gg 1$,
where we neglect the influence of disorder
with the characteristic scattering time $\tau_\text{dis}$.
In the FL case it is known that the presence of disorder has strong influence
on the inelastic relaxation of particles in the diffusive regime $T\tau_\text{dis}\ll 1$.
\cite{Schmid1974,AA1979,Abrahams1981}
However, even in the diffusive regime the tails of the JSD $\propto\omega^{-5/2}$
are preserved for $\omega\tau_\text{dis}\gg 1$,
since they emerge due to scattering with large momentum transfer.

We finish this section with a short discussion of corrections to the results
above due to nonlinearity of the spectrum at high energies $\ep^*\lesssim\Lambda$,
where $\Lambda$ is the cutoff energy.
The nonlinear correction to the dispersion relation reads
$\ep_\lambda(p)-\lambda k \propto k^2 \sin\varphi_k/\Lambda$,
where $\varphi_k$ is the angle of the direction of $\vec{k}$.
The parameter that controls violations of the linear dispersion relation
is therefore $\ep^*/\Lambda$.
Here $\ep^*\sim\omega_{\text{pump}}$ is a characteristic energy.
A positive curvature of the spectrum opens a phase space
for Auger processes (see Appendix~\ref{sec:app_Auger}).
Auger processes thus also contribute to the tail of the JSD.
From a simple estimate (see Appendix~\ref{sec:app_Auger})
we obtain that Auger processes dominate over intraband processes for
$\omega\gtrsim T (T^{1/3}\Lambda^{2/3}/\ep^*)^{2}$.
This region is irrelevant if
$\ep^*\lesssim \Lambda (T/\Lambda)^{5/9}$.
Under this condition the nonlinearity does not modify the tail
of $P(\omega)$.
For room temperature and the cutoff $\Lambda=1\text{eV}$,
even near-infrared to visible light is within the range of validity
of the results of this section.
Since positive curvature only occurs in certain directions,
Auger processes should be even weaker
than in the simple estimate above.
We want to stress that a negative curvature
prevents Auger processes. Negative curvature appears
due to intrinsic band curvature and due to renormalization
of the electron spectrum.

\section{Relaxation cascade: L\'evy flights \label{sec:cascade}}
\noindent
We have seen that the JSD of a high energy electron with energy $\ep_p\gg\max(|\mu|,T)$
in graphene implies an average jump size of the order of either temperature
or chemical potential.
This is in contrast to the FL result where the JSD is flat up to the particle's energy.
In graphene, the excited carriers relax in a cascade, with on average
$\langle n\rangle \sim\ep_p/\langle\omega\rangle$ jumps, where $\langle\dots\rangle$ is the average
according to the JSD. The time scale of the cascade is then $t \sim n/\Gamma$.\cite{SongKoppens2013}

The above conclusion concerns the mean number of steps in the cascade as well
as the average cascade time.
We now discuss the statistics of the random walk modeling the relaxation cascade
in more detail with an emphasis on the fluctuations of the number of cascade steps.

\begin{figure}[h]
	\includegraphics[width=\linewidth]{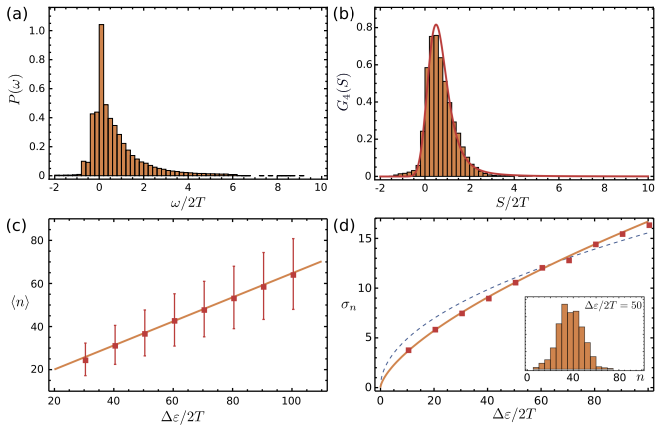}
	\caption{(Color online) (a) Sample of the JSD for $T\gg|\mu|$
	(see Fig.~\ref{sec:cascade_step:fig:P_mu_zero}).
	(b) Sample of the cascade variable $S_n=\omega_1+\dots+\omega_n$
	from the JSD for $n=4$ with a high energy cutoff for the JSD
	given by the particle energy $\ep_p/T=100$.
	The solid line is the stable distribution with $\alpha=3/2$ and $\beta=1$.
	(c) The average number of steps sampled from the JSD as a function
	of the cascade length $\Delta\ep$. The error bars show the typical
	fluctuations $\sigma_n$ of the number of cascade steps.
	(d) The fluctuation $\sigma_n$ as a function of the
	cascade length $\Delta\ep$.
	The solid line is the $\Delta\ep^{2/3}$ law~\eqref{sec:cascade:eq:time_fluctuations}.
	The dashed line illustrates Gaussian fluctuations for comparison.
	The inset shows a typical distribution of cascade steps for $\Delta\ep/2T=50$.
	\label{sec:cascade:fig:levy_sampling}}
\end{figure}

Due to the fact that the JSD exhibits the fat tail $P(\omega)\propto\omega^{-5/2}$,
it does not possess a second moment. Therefore, the fluctuations of the number
of cascade steps should show an unusual behavior.
The particle energy provides a natural cutoff for the JSD, rendering
its variance finite. But on an intermediate scale,
before the electron energy reaches $\max(|\mu|,T)$,
the distribution behaves as if it possessed no finite variance.
This is demonstrated in Fig.~\ref{sec:cascade:fig:levy_sampling}(a)-(b)
by numerical sampling the JSD [Fig.~\ref{sec:cascade:fig:levy_sampling}(a)]
and the cascade $S_n=\omega_1+\dots+\omega_n$
[Fig.~\ref{sec:cascade:fig:levy_sampling}(b)],
where $\omega_i$ are independent and identically distributed.
For not too large $n$, a finite cutoff in the JSD does not change the
distribution of $S_n$ in Fig.~\ref{sec:cascade:fig:levy_sampling}(b).

As a consequence, the large-$n$ limit of the distribution of the cascade
$S_n$ does not approach the normal distribution.
It rather lies in the domain of attraction of an $\alpha$-stable law $G_n(S_n)$.
These are generalized limiting distributions for random processes with stationary
and independent jumps including fat-tailed distributions
as well as the normal distribution ($\alpha=2$).\cite{Feller1971}
Their characteristic function (excluding the case $\alpha=1$ irrelevant for us),
\begin{equation}
	\Phi_n(\alpha,\delta,\beta,c;z) = \ee^{
		\ii z n \delta
		-nc|z|^\alpha(1-\ii\beta\sgn(z)\tan(\alpha\pi/2))
	} \: ,
	\label{sec:cascade:eq:charact_fct}
\end{equation}
is fully parameterized by four parameters.
The index of stability $\alpha=3/2$ follows
from the condition that $G_n(S_n)$ lies in the domain of attraction
of an $\alpha$-stable law,
\begin{equation}
	\int\limits_0^x \dd\omega P(\omega)\omega^2\propto x^{2-\alpha} \: ,
\end{equation}
since the large-$\omega$ asymptotic of the JSD, $P(\omega)$, is given by
\begin{equation}
	P(\omega) T/\Gamma \simeq c \: (\omega/2T)^{-5/2} \: .
	\label{sec:cascade:eq:asymptotic}
\end{equation}
The scale parameter $c$ is obtained from Eqs.~\eqref{sec:app_BE:eq:asymptotics_3}
and \eqref{sec:app_BE:eq:asymptotics_6}.
It will be related to the anomalous diffusion constant in Sec.~\ref{sec:FFPE},
Eq.~\eqref{sec:FFPE:eq:anomalous_diffusion_const}.
The skewness $\beta=1$ in the case of graphene, rendering the distribution single
sided - the electron loses energy in the cascade.
The location parameter $\delta=\langle\omega\rangle$.
For $|\mu|\gg T$ we have $\delta\sim\alpha_g|\mu|$
whereas $\delta\sim \alpha_g T$ for $T\gg|\mu|$.

The random variable $Y=S_n-n\delta$, describing the fluctuations of the cascade,
obeys a strictly stable distribution.
The random motion on top of the drift during the relaxation processes
is thus not the standard Brownian motion but is rather superdiffusive containing
long jumps. The associated statistics
serves as a fingerprint of the EEI in graphene.

We discuss three important consequences:

(i) The relaxation rate $\gamma_c$ of the entire cascade is given by the rate $\Gamma$
divided by the average number of steps. The latter is given by $\ep_p/\langle\omega\rangle$.
Thus we obtain
\begin{equation}
	\gamma_c \sim
	\alpha_g^2
	\begin{cases}
		\mu^2/\ep_p \: , &  \: |\mu|\gg T \\
		T^2/\ep_p \:  ,& \: T\gg|\mu|
	\end{cases}
	\: .
	\label{sec:cascade:eq:cascaderate}
\end{equation}

\begin{figure}[h]
	\includegraphics[width=\linewidth]{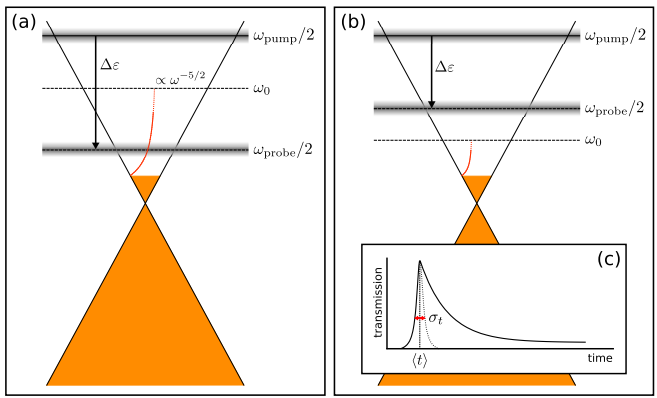}
	\caption{(Color online) Pump-probe setup for
	(a) $\omega_\text{probe}<\omega_0$: The probe measures mostly
	the secondary particles which are created with the probability
	$P(\omega)\propto\omega^{-5/2}$;
	(b) $\omega_\text{probe}>\omega_0$: The density of secondary particles
	is negligible and the situation is suitable for studying the cascade time
	and its fluctuations depending on the length of the cascade
	$\Delta\ep=(\omega_\text{pump}-\omega_\text{probe})/2$.
	(c) The fluctuations~\eqref{sec:cascade:eq:time_fluctuations}
	determine the width of the rise time in the measured change of the transmission
	[see also the inset of Fig.~\ref{sec:cascade:fig:levy_sampling}(d)].
	\label{sec:cascade:fig:experiment}}
\end{figure}

(ii) Second, the high-energy tail of the JSD $P(\omega^*)$, $\omega^*\gg \max(|\mu|,T)$, gives also the
probability density for a secondary electron or hole to be created in the energy interval $\omega^*\lesssim|\varepsilon|\lesssim\omega^*+\max(|\mu|,T)$. 
More precisely, in the case $\mu\gg T$ ($-\mu\gg T$) only hot electrons (holes) are created
with probability density $P(\omega^*)$, while in the case $T\gg |\mu|$ electrons and holes are created
with equal probability $P(\omega^*)/2$. 
Using $P(\omega^*)\ll P(\langle \omega \rangle)$, the probability to create a secondary electron at energy $\ep\sim\omega^*$ during the entire cascade
is then given (up to the factor 1/2) by $P(\omega^*)\ep_p/\langle \omega \rangle$.
We conclude that the energy scale
\begin{equation}
	\omega_0 \sim
	\begin{cases}
		T(\ep_p/\alpha_g|\mu|)^{2/5} \: ,& \: |\mu|\gg T \\
		T(\ep_p/\alpha_g T)^{2/5} \: ,& \: T\gg|\mu|
	\end{cases}
	\: ,
	\label{sec:cascade:eq:omega_0}
\end{equation}
separates the regions where the density of downstream particles
is smaller ($\omega^*<\omega_0$) and larger ($\omega^*>\omega_0$) than the density of secondary particles,
see Fig.~\ref{sec:cascade:fig:experiment}(a)-(b).
In the former region the distribution function should show traces of the tail of the JSD
accordingly [Fig.~\ref{sec:cascade:fig:experiment}(a)].
\footnote{
In fact secondary electrons generated during the cascade will also relax.
The account for this relaxation requires the full solution of the kinetic equation
which is beyond the scope of this work.
}

(iii) The third consequence concerns the scaling behavior of fluctuations of the
cascade time - the first passage time of the L\'evy process on the finite distance $\Delta\ep$ in the energy space -
which is directly related to the random variable $Y$.
The distance $\Delta\ep$ can be for instance given by $\Delta\ep=(\omega_\text{pump}-\omega_\text{probe})/2$, the difference between the
excitation and probing frequency, see Fig.~\ref{sec:cascade:fig:experiment}.
We use the scaling of L\'evy stable distributions,
\begin{equation}
	G_n(S_n) = n^{-1/\alpha} G_1(Y/n^{1/\alpha})|_{\delta=0} \: ,
	\label{sec:cascade:eq:scaling}
\end{equation}
that follows from Eq.~\eqref{sec:cascade:eq:charact_fct} and obtain
\begin{equation}
	\langle Y^2 \rangle
	\sim
	\Delta\ep^{2/\alpha} T^{2(\alpha-1)/\alpha}\: .
	\label{sec:cascade:eq:fluct_relation}
\end{equation}
The mean square fluctuation of the number of steps is then given by 
$\sigma_n^2 = \langle n^2 \rangle-\langle n \rangle^2= \delta^{-2}\langle Y^2 \rangle$ while
the fluctuation of the cascade time 
\begin{equation}
	\sigma_t = \Gamma^{-1}\sigma_n = T^{(\alpha-1)/\alpha}\Delta\ep^{1/\alpha}/\Gamma\langle\omega\rangle\: .
	\label{sec:cascade:eq:time_fluctuations}
\end{equation}
Using Eq.~\eqref{sec:cascade_step:eq:Gamma_mu_zero} and \eqref{sec:cascade_step:eq:Gamma_mu_finite}
in Eq.~\eqref{sec:cascade:eq:time_fluctuations} we obtain,
\begin{equation}
	\sigma_t \sim \left( \frac{\Delta\ep}{T} \right)^{1/\alpha}
	\begin{cases}
		T/\mu^2 \: ,&  \: |\mu|\gg T \\
		T^{-1} \: ,& \: T\gg|\mu|
	\end{cases}
	\: .
	\label{sec:cascade:eq:time_fluct_expl}
\end{equation}
Both for $|\mu|\gg T$ and for $T\gg|\mu|$  we find a nontrivial dependence on $\sigma_t(T)$
determined by the index of stability $\alpha$.
Since $\alpha=3/2$ in our case, the fluctuations increase
$\propto T^{1/3}$ at $T\ll|\mu|$ and decrease $\propto T^{-5/3}$ at $T\gg|\mu|$.

The dependence of the fluctuations in the number of cascade steps $n$
on the length of the cascade $\Delta\ep$ is demonstrated
in Figs.~\ref{sec:cascade:fig:levy_sampling}(c)-(d).
Here the cascade is simulated by generating a sequence of steps from the JSD
until the cascade length $\Delta\ep$ is reached.
The average number of steps $\langle n\rangle$
in Fig.~\ref{sec:cascade:fig:levy_sampling}(c)
scales linearly with the cascade length $\Delta\ep$.
On the other hand, the fluctuations of the number of steps $\sigma_n$
in Fig.~\ref{sec:cascade:fig:levy_sampling}(d)
obey the relation~\eqref{sec:cascade:eq:time_fluctuations}.

The exponent of $\Delta\ep$ in the fluctuations $\sigma_t$,
Eq.~\eqref{sec:cascade:eq:time_fluctuations}, is known as the Hurst
exponent $H=1/\alpha$.\cite{Hurst1951,FogedbyHurst1992}
It is related to the fractal dimension of the random walk $D_f=2-H=4/3$.
\cite{fractalsFeder1988}
The fractal nature of the relaxation cascade in graphene can be understood
in terms of a fast one-dimensional backbone of forward scattering
augmented by other less efficient channels in the 2D momentum space,
similar to the emergence of fractal dimensions in networks.

\section{Fractional kinetics and transient change in transmission \label{sec:FFPE}}
\noindent
In this section we will calculate the transient differential transmission of
a graphene sample after laser excitation.
As in the previous sections we assume that the density of high energy electrons
is much lower than the density of thermal electrons and we can neglect the mutual
interaction of the excited carriers.
Second, we calculate the isotropic part of the distribution function
at high energies $\ep>\omega_0$ [see Fig.~\ref{sec:cascade:fig:experiment}(b)],
thus we can neglect secondary electrons.
Furthermore we neglect the exponential tail of the thermal electrons since
$\ep\gg\max(|\mu|,T)$. Therefore the isotropic part of the
transient distribution function will be given by the
distribution of downstream electrons, denoted $F(\ep,t)$.
\subsection{Fractional kinetics}
\noindent
In the previous section we showed that the statistics of the relaxation dynamics
is given by L\'evy flights. In terms of the distribution function the relaxation
will be described by the fractional Fokker-Planck equation (FFPE),\cite{Metzler1999}
\begin{equation}
	\partial_t W(\ep,t) =
	\Gamma\langle\omega\rangle \: \partial_\ep W(\ep,t)
	+ D \nabla_{(\beta)}^\alpha W(\ep,t)
	\: .
	\label{sec:FFPE:eq:FFPE}
\end{equation}
Here $W(\ep,t)$ with $W(\ep,t=0)=\delta(\ep)$ is the propagator of the FFPE
which will be given below.
We also introduced the Riesz-Feller fractional derivative,\cite{Mainardi2005}
which is defined by its Fourier transform,
\begin{equation}
	\nabla_{(\beta)}^\alpha
	f(\ep)
	= \int \frac{\dd z}{2\pi} \:
		\ln[\Phi_1(\alpha,0,\beta,1;z)] f(z)\ee^{\ii z \ep} \: ,
	\label{sec:FFPE:eq:Riesz-Feller_fract_der}
\end{equation}
where $\Phi_1$ is the characteristic function of the
underlying stochastic process. In our case it is a L\'evy $\alpha$-stable law
with $\alpha=3/2$ and $\beta=1$, see Eq.~\eqref{sec:cascade:eq:charact_fct}.
In the FFPE~\eqref{sec:FFPE:eq:FFPE} we also introduced the average
energy loss rate $\Gamma\langle\omega\rangle$
and the anomalous diffusion constant $D=\Gamma c$,
where $c$ is the scale paramter of the L\'evy process,
see Eqs.~\eqref{sec:cascade:eq:charact_fct} and \eqref{sec:cascade:eq:asymptotic}.
From these formulas we obtain
\begin{equation}
	D =
	\frac{2^\alpha 128 \sqrt{2\pi}}{N/4} \: T^{\alpha+1}
	\: .
	\label{sec:FFPE:eq:anomalous_diffusion_const}
\end{equation}

The emergence of the fractional kinetics expressed by the FFPE~\eqref{sec:FFPE:eq:FFPE}
can be understood on the basis of a Langevin-type rate equation for the electron energy,
\begin{equation}
	\partial_t \ep(t) = -\Gamma\langle\omega\rangle +\eta(t)
	\: ,
	\label{sec:FFPE:eq:Langevin}
\end{equation}
where $\eta(t)$ is a random variable which is distributed
according to an $\alpha$-stable law and describes the interaction
of the high energy electron with the bath of thermal electrons.

The general solution $F(\ep,t)$ of the FFPE with initial conditions
$F(\ep,t=0)=f(\ep)$ is obtained with the propagator according to
\begin{equation}
	F(\ep,t) = \int \dd \ep' \: W(\ep-\ep',t)f(\ep') \: .
	\label{sec:FFPE:eq:P_x}
\end{equation}
In our case we choose the initial probability density to be
\begin{equation}
	f(\ep) = n_0 \delta(\ep-\omega_\text{pump}/2) \: .
	\label{sec:FFPE:eq:inital_cond}
\end{equation}
Here $n_0$ is the integrated flux density of the pump pulse.
\footnote{
For a Gaussian initial density
$f(\ep) = (n_0/\delta\sqrt{2\pi}) \exp[(\ep-\omega_\text{pump}/2)^2/2\delta^2]$
with width $\delta$, our results remain valid for large times $t\gg\delta^\alpha/D$,
when the initial condition is washed out and the form of the probability
density is determined by diffusion.
}
We have $F(\ep,t)= n_0 W(\ep_t,t)$, where 
\begin{equation}
	\ep_t=\ep-\omega_\text{pump}/2+\Gamma\langle\omega\rangle t
	\: ,
	\label{sec:FFPE:eq:running_energy}
\end{equation}
is the running energy.
The propagator $W(\ep,t)$ and thus the solution $F(\ep,t)$
in our case of $\alpha=3/2$ and $\beta=1$ can be calculated explicitely.
\footnote{The propagator can be written for arbitrary $\alpha$ and
$\beta$ in terms of the Fox H-function.}
We obtain
\begin{equation}
	W(\ep,t)=\frac{\pi T}{\alpha}(D t)^{-1/\alpha}K(s)
	\label{sec:FFPE:eq:sol_W}
\end{equation}
for the propagator in terms of the dimensionless variable
\begin{equation}
	s=\ep_t/ (D t)^{1/\alpha}
	\: .
	\label{sec:FFPE:eq:similarity_var}
\end{equation}
In Eq.~\eqref{sec:FFPE:eq:sol_W} the function $K(s)$ is given by,
\begin{equation}
	K(s) = -\ee^{\frac{s^{3}}{27}}
	\left[
		\sqrt[3]{2/3} \: s \text{Ai}\left(\frac{s^2}{\sqrt[3]{486}}\right)
		+\sqrt[3]{12} \: \text{Ai}'\left(\frac{s^2}{\sqrt[3]{486}}\right)
	\right]
	\: .
	\label{sec:FFPE:eq:K}
\end{equation}
Here $\text{Ai}(z)$ is the Airy function and $\text{Ai}'(z)$ its derivative.
In particular, $W$ has the following asymptotics for large times,
\begin{equation}
	W(\ep,t)
	\simeq
	\frac{T D t}{\sqrt{2\pi}\alpha} \: |\ep-\ep_0+\Gamma\langle\omega\rangle t|^{-(\alpha+1)}
	\: .
	\label{sec:FFPE:eq:W_asymptotics}
\end{equation}
Using Eq.~\eqref{sec:FFPE:eq:anomalous_diffusion_const} and
the results from Sec.~\ref{sec:cascade_step} we obtain,
\begin{equation}
	W(\ep,t)
	\sim
	t^{-\alpha}
	\begin{cases}
		T(T/\mu^2)^{\alpha+1} \: ,&  \: |\mu|\gg T \\
		T^{-\alpha} \: ,& \: T\gg|\mu|
	\end{cases}
	\: .
	\label{sec:FFPE:eq:W_asymptotics}
\end{equation}
We see that the tail of F for large times but fixed $\ep$
is proportional to $t^{-3/2}$
and scales as $T^{-3/2}$ for $T\gg|\mu|$
and as $T(T/\mu^2)^{5/2}$ for $|\mu|\gg T$.

\begin{figure}[h]
	\includegraphics[width=\linewidth]{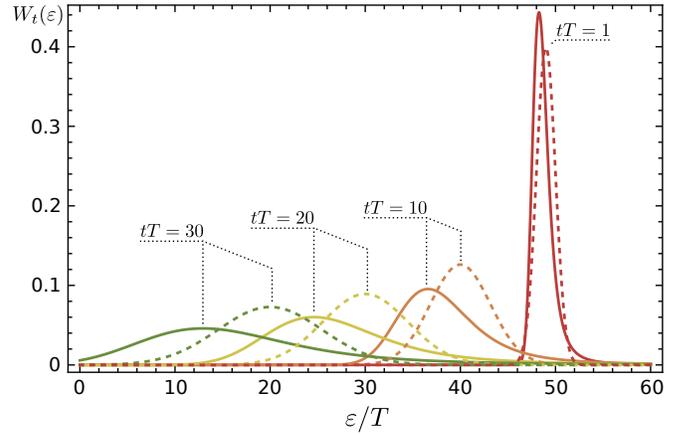}
	\caption{(Color online) The solution
	[see Eqs.~\eqref{sec:FFPE:eq:sol_W}-\eqref{sec:FFPE:eq:K}]
	of the FFPE~\eqref{sec:FFPE:eq:FFPE} (solid line)
	as a function of energy in comparison to the result obtained for
	Gaussian diffusion (dashed line) for different times.
	\label{sec:FFPE:fig:wave_packet}
	}
\end{figure}

The evolution of the probability distribution $W(\ep,t)$ due to the fractional
kinetics is illustrated in Fig.~\ref{sec:FFPE:fig:wave_packet}.
The solid line depicts the solution of the FFPE~\eqref{sec:FFPE:eq:FFPE},
given by the Eqs.~\eqref{sec:FFPE:eq:sol_W}-\eqref{sec:FFPE:eq:K},
while the dashed lines show the Gaussian solution of the usual Fokker-Planck
equation.
The fractional kinetics leads to a strong asymmetry,
compared to the Gaussian drift-diffusion,
since the fluctuations in the underlying L\'evy process are single sided,
i.e. $\beta=1$ in Eq.~\eqref{sec:cascade:eq:charact_fct} and \eqref{sec:FFPE:eq:FFPE}.
\subsection{Transient change in transmission}
\noindent
We outline the consequences of the fractional kinetics
for the transient differential transmission of the sample.
The latter is determined by the change in the dynamic conductivity which is given by,
\begin{equation}
	\Delta\sigma(t)/\sigma_0
	= -\left[ F(\omega_\text{probe}/2,t)-F(-\omega_\text{probe}/2,t)\right]
	\: .
	\label{sec:FFPE:eq:sigma}
\end{equation}
Given the particle hole symmetry of the correction to the distribution
function at high energies, i.e. $F(-\ep,t)=-F(\ep,t)$, we finally have
for the relative differential transmission
\begin{equation}
	\frac{\Delta T(t)}{T_0}
	= 2 n_0 W(\omega_\text{probe}/2,t) \: ,
	\label{sec:FFPE:eq:DeltaT}
\end{equation}
where $n_0$ is the integrated flux density.

\begin{figure}[h]
	\includegraphics[width=\linewidth]{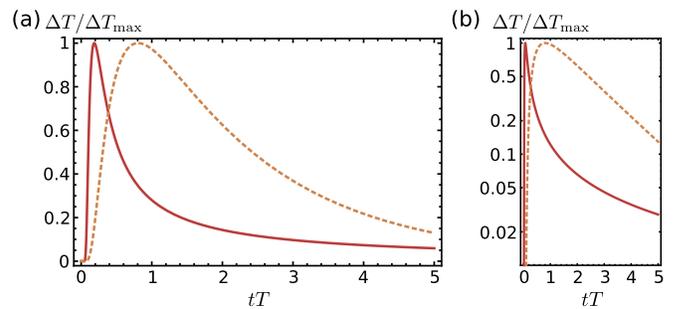}
	\caption{(Color online) The normalized differential transmission
	$\Delta T/\Delta T_\text{max}$ as a function of the dimensionless time $tT$.
	Here $\Delta T_\text{max}$ denotes the maximum value of $\Delta T$.
	Figure (b) shows the results on a logarithmic scale.
	The solid curves are calculated according to Eq.~\eqref{sec:FFPE:eq:DeltaT}
	and Eqs.~\eqref{sec:FFPE:eq:sol_W} and \eqref{sec:FFPE:eq:K}, for
	$\Delta\ep/T=(\omega_\text{pump}-\omega_\text{probe})/T=25$
	and $\Gamma\langle\omega\rangle/T^2=20$ as well as $D/T^{\alpha+1}$
	from Eq.~\eqref{sec:FFPE:eq:anomalous_diffusion_const}.
	The dashed lines in (a) and (b) illustrate the result for usual
	diffusion in comparison to the fractional kinetics (solid line).
	\label{sec:FFPE:fig:transient_transmission}
	}
\end{figure}

The behavior of $\Delta T$ as a function of time,
Eq.~\eqref{sec:FFPE:eq:DeltaT}, is illustrated in
Fig.~\ref{sec:FFPE:fig:transient_transmission}.
The solid line depicts the result~\eqref{sec:FFPE:eq:sol_W} due to
the fractional kinetics in graphene, while the dashed line is the expected result
for conventional Gaussian drift-diffusion.
We see that the diffusion in the case of L\'evy flights (solid line) is stronger due
to the fact that the $\alpha$-stable law is single sided, i.e. $\beta=1$.
Therefore fluctuations enhance the drift in energy space,
see also Fig.~\ref{sec:FFPE:fig:wave_packet}.
Furthermore the transient differential transmission shows powerlaw behavior
with time and temperature according to Eq.~\eqref{sec:FFPE:eq:W_asymptotics},
instead of exponential decay in the case of usual diffusion
[see Fig.~\ref{sec:FFPE:fig:transient_transmission}(b)].

\section{Conclusion \label{sec:conclusion}}
\noindent
We have provided an analysis of the relaxation cascade of photoexcited
electrons in graphene at finite temperature.
We calculated the relaxation rates of high-energy electrons in the case of doped
as well as undoped graphene. We find $\Gamma\sim\alpha_g\max(|\mu|,T)$,
which distinguishes graphene from the FL.
The $\alpha_g$ dependence deviates distinctively from the golden rule result $\propto\alpha_g^2$
and is due to the peculiar screening in graphene.\cite{Schuett2011}
Furthermore the rates are independent of the particle energy $\ep_p$.
The entire relaxation cascade is determined by the distribution of the
transfered energy in a single jump.
This jump-size distribution (JSD) exhibits logarithmic divergencies
at small energy transfer due to resonant forward scattering
which is very pronounced in graphene having truly linear spectrum.
Specifically, we find $P(\omega)\sim \ln\alpha_g T/|\omega|$ for $|\mu|\ll T$
and small frequencies $|\omega|\ll\alpha_g T$ which crosses over into
the usual FL result $P(\omega)\sim(T/|\mu|)\ln\alpha_g|\mu/\omega|$
at $|\mu|\gg T$.
Remarkably, the JSD exhibits fat tails that fall off as $(\omega/T)^{-5/2}$
at large frequencies $\omega>\text{max}(2|\mu|,2T)$ for both $|\mu|\gg T$ and $T\gg|\mu|$.

Owing to the fat-tailed JSD, the relaxation cascade is described by
an $\alpha$-stable distribution with a mean drift determined by either $T$ or $|\mu|$:
The fluctuations on top of the drift is described by L\'evy flights with
index of stability $\alpha=3/2$.
As a consequence, the fluctuations $\sigma_t$ of the cascade time $t$
exhibit characteristic scaling relations with the frequency $\omega_\text{pump}\gg\omega_\text{probe}$
of the pump pulse, $\sigma_t\sim \omega_\text{pump}^{1/\alpha}$,
as well as temperature. Specifically, $\sigma_t\sim T^{1/3}$ for $|\mu|\gg T$
and $\sigma_t\sim T^{-5/3}$ for $T\gg|\mu|$.
These scaling relations serve a clear imprint of the forward scattering resonance
and related fractal nature of the relaxation cascade in graphene. 
The observed \cite{TielrooijKoppens2013} variation of the average cascade time with $\omega_\text{pump}$ 
is consistent with theoretical predictions for the energy drift made in Ref.~\onlinecite{SongKoppens2013}
for the regime $|\mu|\gg T$. Using the experimental setup similar to that used in Refs.~\onlinecite{Brida2013, TielrooijKoppens2013},
it should be possible to detect the traces of the Levy flights as well. Specifically, the width of the rise time in the measured change in 
transmission as depicted in Fig.~\ref{sec:cascade:fig:experiment}(c) provides a direct measure of the fluctuation 
of the cascade time  (\ref{sec:cascade:eq:time_fluct_expl}) [see also Fig.~\ref{sec:cascade:fig:levy_sampling}(d)].

Furthermore, the JSD is the distribution of the created electron-hole pairs during
the cascade. We find that within the energy window $\max(|\mu|,T)\ll\ep<\omega_0$
a significant amount of secondary electrons are created according to
$P(\omega)\simeq\omega^{-5/2}$.
We find $\omega_0\sim T(\omega_\text{pump}/T)^{2/5}$ for $T\gg|\mu|$ and
$\omega_0\sim T(\omega_\text{pump}/\mu)^{2/5}$ for $|\mu|\gg T$.
Probes in the mentioned energy interval should also reveal the tail of the JSD.

We predict the time evolution of the differential change in transmission
in the presence of electron electron intercations.
The transmission is directly measured in pump probe experiments
and we obtain an analytical expression for the differential transmission
from a fractional Fokker-Planck equation.
The latter is suited to capture the fractional kinetics emerging
from the L\'evy flight statistics of the relaxation process.

The results of this work extend the study of relaxation dynamics of 
thermal electrons in graphene\cite{Schuett2011} to the case of high energy electrons
also at finite chemical potential
and should be relevant for future studies of the nonequilibrium steady 
states in irradiated graphene. This prospect includes the question of thermalization 
in driven graphene, the possibility of a population inversion\cite{Rhizzi2007,PopInv2012}
as well as frequency conversion.\cite{Bonaccorso2010}
It should also be interesting to extend it to the non-linear regime of pumping
where saturation effects become important.
All these questions necessitate the full solution of the kinetic equation.
In this context the present work sheds new light on the unique character of the interaction
in graphene that controls the formation of such nonequilibrium states,
that might also be probed in future experiments.

\begin{acknowledgments}
\noindent
We would like to thank I. Gornyi, B. Jeevanesan, M. Sch\"utt and C. Seiler for useful discussions.
Furthermore we are indebted to F. Koppens and K.-J. Tielrooij for discussions
and providing insights into experiments.

This work was supported by DFG in the framework of the SPP 1459 and by BMBF.

\end{acknowledgments}


%

\appendix
\onecolumngrid
\section{Calculation of the relaxation rate and the JSD from the Boltzmann equation \label{sec:app_BE}}
\noindent
In this section we derive Eq.~\eqref{sec:cascade_step:eq:P_definition} for the JSD and the
relaxation rates from the main text.
We start from a generic fermionic collision integral
\begin{equation}
	\textit{St}[f(\lambda,\vec{p})]
	= 
	\sum_{\lambda_3} \int\frac{\dd^2 p_3}{(2\pi)^2} \: \bigg\{
	W(\lambda,\vec{p}|\lambda_3,\vec{p}_3) f_{\lambda_3}(\vec{p}_3)
	[1-f_{\lambda}(\vec{p})]
	- W(\lambda_3,\vec{p}_3|\lambda,\vec{p})f_{\lambda}(\vec{p})[1-f_{\lambda_3}(\vec{p}_3)] \bigg\} \: ,
	\label{sec:app_BE:eq:coll_integral}
\end{equation}
where the transition rates for the Coulomb interaction
\begin{equation}
	\begin{split}
		& W(\vec{p}_2,\lambda_2|\vec{p}_1,\lambda_1) = (2\pi)^{-1} \sum_{\lambda_3,\lambda_4}
		\int \dd\vec{p}_{3,4} \int \dd\vec{q}\dd\omega
		\delta(\lambda_2 p_2+\omega-\lambda_1 p_1) \delta(\lambda_4 p_4 -\omega-\lambda_3 p_3) \\
		& \times \delta(\vec{p}_2+\vec{q}-\vec{p}_1) \delta(\vec{p}_4-\vec{q}-\vec{p}_3) 
		K(\vec{q},\omega,\{\lambda_i\},\{\vec{v}_i\}) f(\vec{p}_3,\lambda_3)[1-f(\vec{p}_4,\lambda_4)] \: .
	\end{split}
	\label{sec:app_BE:eq:transition_rates}
\end{equation}
Here the interaction kernel
\begin{equation}
	K(q,\omega,\{\lambda_i\},\{\mb{v}_i\}) = N |V(\omega,q)|^2 \Theta_{1,2} \Theta_{3,4} \: ,
	\label{sec:app_BE:eq:coll_kernel}
\end{equation}
contains the RPA screened Coulomb matrix element (see App.~\ref{sec:app_RPA})
\begin{equation}
	|V(\omega,q)|^2 = 
	\frac{4\pi^2\alpha_g^2}{(q+2\pi\alpha_g N \text{Re}\Pi)^2+(2\pi\alpha_g N\text{Im}\Pi)^2} \: ,
	\label{sec:app_BE:eq:P_2}
\end{equation}
as well as the Dirac factors ($\vec{v}_i=\lambda_i\vec{k}_i/k_i$) $ \Theta_{1,2} = (1+\vec{v}_1\cdot\vec{v}_2)/2$.
Upon inserting the ansatz~\eqref{sec:cascade_step:eq:ansatz_f}
into the collision integral~\eqref{sec:app_BE:eq:coll_integral},
we obtain the explicit expression for the relaxation rate
\begin{equation}
 	\Gamma = \sum_{\lambda_1}\int\frac{\dd^2 k}{(2\pi)^2}
	\: W_0(\lambda_1,\vec{k}|+1,\vec{p})[1-f_{\lambda_1}(\vec{k})]
	+ W_0(+1,\vec{p}|\lambda_1,\vec{k})f_{\lambda_1}(\vec{k})
	\: .
	\label{sec:app_BE:eq:Gamma}
\end{equation}
For $\omega<\ep_p$, where interband processes are forbidden,
the second term in Eq.~\eqref{sec:app_BE:eq:Gamma} can be dropped.
Using Eqs.~\eqref{sec:app_BE:eq:transition_rates}-\eqref{sec:app_BE:eq:coll_kernel} we then obtain
\begin{equation}
	\begin{split}
		& \Gamma = (2\pi)^2 \sum_{\lambda_{1,3,4}}
		\int\frac{\dd^2q}{(2\pi)^2}\frac{\dd\omega}{2\pi}\frac{\dd^2 k_4}{(2\pi)^2} \:
		\delta(\lambda_2|\vec{p}-\vec{q}|+\omega-p)\delta(\lambda_4 k_4-\omega-\lambda_3|\vec{k}_4-\vec{q}|) \\
		& \times N |V_{RPA}(\omega,q)|^2 \:\Theta_{1,2}\big\vert_{1=(\lambda_1,\vec{p}-\vec{q})} \: 
		\Theta_{3,4}\big\vert_{3=(\lambda_3,\vec{k}_4-\vec{q})}
		\: f_T(\lambda_4 k_4-\omega)[1-f_T(\lambda_4 k_4)] \: .
	\end{split}
	\label{sec:app_BE:eq:Gamma}
\end{equation}
Next we perform the angular integration in the integrals over $\vec{k}_4$ and $\vec{q}$.
The arising functional determinants are ($\lambda=+1$),
\begin{align}
	& \left| \frac{\partial}{\partial\varphi_q}\lambda_2 |\vec{p}-\vec{q}| \right|
	= \frac{p q |\sin(\varphi_q-\varphi_p)|}{|\vec{p}+\vec{q}|}
	= \frac{\sqrt{q^2-\omega^2}[(\omega-2\lambda p)^2-q^2]^{1/2}}{2|\lambda p - \omega|} \: ,
	\label{sec:app_BE:eq:funct_det_1} \\
	& \left| \frac{\partial}{\partial\varphi_4}\lambda_3 |\vec{k}_4-\vec{q}| \right|
	= \frac{k_4 q |\sin(\varphi_4-\varphi_q)|}{|\vec{k}_4-\vec{q}|}
	= \frac{\sqrt{q^2-\omega^2}[(\omega-2\lambda_4 k_4)^2-q^2]^{1/2}}{2|\lambda_4 k_4 - \omega|} 
	\: .
	\label{sec:app_BE:eq:funct_det_2}
\end{align}
The corresponding Dirac factors are ($\lambda_1=\lambda=+1$)
\begin{align}
	& \Theta_{1,2}
	=\frac{1}{2}\left(1+\frac{\lambda_1\lambda_2\vec{k}_1\cdot(\vec{p}-\vec{q})}{k_1|\vec{p}-\vec{q}|}\right)
	=\frac{|(\omega-2\lambda p)^2 - q^2|}{4 p|\lambda p -\omega|} \: ,
	\label{sec:app_BE:eq:DiracFact_1} \\
	& \Theta_{3,4}
	=\frac{1}{2}\left(1+\frac{\lambda_2\lambda_3\vec{k}_4\cdot(\vec{k}_4-\vec{q})}{k_4|\vec{k}_4-\vec{q}|}\right)
	=\frac{|(\omega-2\lambda_4 k_4)^2 - q^2|}{4 k_4|\lambda_4 k_4 -\omega|} \: .
	\label{sec:app_BE:eq:DiracFact_2} 
\end{align}
\subsection{The JSD $P(\omega)$}
\noindent
Putting together Eqs.~\eqref{sec:app_BE:eq:Gamma}-\eqref{sec:app_BE:eq:DiracFact_2} we finally obtain the JSD
\begin{equation}
	P(\omega) = \int_0^\infty \dd q \:
	\frac{q\:\text{Re}\sqrt{\sgn(q^2-\omega^2)[(\omega-2\lambda p)^2-q^2]}}{2p}
	\: \frac{N|V(\omega,q)|^2}{|q^2-\omega^2|} \: \mathcal{K}(\omega,q) \: .
	\label{sec:app_BE:eq:P_1}
\end{equation}
Here the kinetic kernel is given by Eq.~\eqref{app:kinKernel:eq:kin_Kernel}.
If we assume $p\gg\omega,q$ we obtain the result~\eqref{sec:cascade_step:eq:P_definition} stated in the main text,
\begin{equation}
	P(\omega) = \int_{|\omega|}^\infty \dd q \: q
	\: \frac{N|V(\omega,q)|^2}{|q^2-\omega^2|} \: \mathcal{K}(\omega,q) \: .
\end{equation}
In the following we use the dimensionless variables $\Omega=\omega/2T$,
$Q=q/2T$, $\beta=\omega/q$ and $\tilde\mu=\mu/T$.
Using the asymptotics from App.~\ref{sec:app_RPA} and \ref{sec:app_K} we obtain
limiting expressions for the JSD $P(\Omega)$ presented below.
\subsubsection{The limit $T\gg|\mu|$ for $|\Omega|<1$}
\noindent
The contribution for small momentum transfer ($Q<1$) reads,
\begin{equation}
	P(\Omega)\big|_{Q<1} =
	4\ln 2 \: \alpha_g^2 \pi^2 N \ee^\Omega \int_{|\Omega|}^{1} \dd Q \:
	\frac{Q}{|Q^2-\Omega^2|(Q+\alpha_g N \ln 2)^2+(\alpha_g N \ln 2 \Omega)^2}
	= \frac{4\pi^2}{N\ln2} \ln\frac{\alpha_g N \ln2}{|\Omega| }\: .
	\label{sec:app_BE:eq:asymptotics_1}
\end{equation}
Here the last equality is valid for $|\Omega|<\alpha_g N \ln 2$.
The contribution to the JSD with large momentum transfer ($Q>1$) for frequencies $|\Omega|<1$ is,
\begin{equation}
	P(\Omega)\big|_{Q>1} =
	2 \alpha_g^2 \pi^2 N \ee^\Omega \int_{1}^{\infty} \dd Q \:
	\frac{\sqrt{2\pi}Q^{3/2}\ee^{-Q}}
	{(\sqrt{Q^2-\Omega^2}Q+\alpha_g\pi N Q^2/16)^2+(\alpha_g\pi N \ee^{-Q}\sqrt{Q/2\pi} )^2}
	\: .
	\label{sec:app_BE:eq:asymptotics_2}
\end{equation}
The latter can be neglected for $|\Omega|<\alpha_g$.
\subsubsection{The limit $T\gg\mu|$ for $|\Omega|>1$ ($Q>1$)}
\noindent
For $|\Omega|>1$, where only intraband transitions with $Q>1$ are possible, the JSD reads,
\begin{equation}
	P(\Omega) =
	2 \alpha_g^2 \pi^2 N \ee^\Omega \int_{|\Omega|}^{\infty} \dd Q \:
	\frac{\sqrt{2\pi}Q^{3/2}\ee^{-Q}}
	{(\sqrt{Q^2-\Omega^2}Q+\alpha_g\pi N Q^2/16)^2+(\alpha_g\pi N \ee^{-Q}\sqrt{Q/2\pi} )^2}
	\simeq \frac{2^9\sqrt{2\pi}}{N} \: |\Omega|^{-5/2} \: ,
	\label{sec:app_BE:eq:asymptotics_3}
\end{equation}
where the asymptotics is valid for $|\Omega|\gg 1$.
\subsubsection{The limit $|\mu|\gg T$ for $|\Omega|<|\tilde\mu|$ ($Q<|\tilde\mu|$)}
\noindent
\begin{equation}
	P(\Omega) = 4 \alpha_g^2 \pi^2 N  \: \Omega |\tilde\mu|  (1+\coth(\Omega))
	\int_{|\Omega|}^{|\tilde\mu|} \dd Q \:
	\frac{Q}{(Q^2-\Omega^2)(Q+\alpha_g N |\tilde\mu|/2)^2+(\alpha_g N \tilde\mu \Omega/2)^2} \: .
	\label{sec:app_BE:eq:asymptotics_4}
\end{equation}
Equation~\eqref{sec:app_BE:eq:asymptotics_4} can be integrated analytically,
similar to Eq.~\eqref{sec:app_BE:eq:asymptotics_1}, yielding a lengthy expression.
For brevity we give the limit for $|\Omega|\ll\alpha_g |\mu|$,
\begin{equation}
	P(\Omega) \simeq \frac{1}{32\pi}\frac{\ln\alpha_g N|\mu/2\Omega|}{|\mu|^2} \: .
	\label{sec:app_BE:eq:asymptotics_5}
\end{equation}
\subsubsection{The limit $|\mu|\gg T$ for $|\Omega|>|\tilde\mu|$ ($Q>|\tilde\mu|$)}
\noindent
As in the case $|\tilde\mu|\ll 1$, here for $|\Omega|>|\tilde\mu|$ the JSD is
determined by scattering with large momentum transfer,
\begin{equation}
	P(\Omega) = 2 \alpha_g^2 \pi^2 N  \: \ee^{+\Omega} \:
	\int_{|\Omega|}^{\infty} \dd Q \: \frac{\sqrt{2\pi}Q^{3/2}\ee^{-Q}}{(\sqrt{Q^2-\Omega^2}Q+\alpha_g \pi N Q^2 /16)^2}
	\simeq \frac{2^9\sqrt{2\pi}}{N} \: |\Omega|^{-5/2} \: .
	\label{sec:app_BE:eq:asymptotics_6}
\end{equation}
\subsection{The relaxation rate $\Gamma$}
\noindent
\subsubsection{The limit $T\gg|\mu|$}
\noindent
We first calculate the relaxation rate for $T\gg|\mu|$.
We find that the contribution from the region with $Q>1$ is of order $\alpha_g^2$,
whereas $|\Omega|<1$ yields the leading contribution $\propto\alpha_g$:
\begin{equation}
	\Gamma/2T = \int_{0}^{\alpha N \ln 2}\dd\Omega \: P(\Omega)\big|_{Q<1}
	+ \int_{\alpha N \ln 2}^{1}\dd\Omega \: P(\Omega)\big|_{Q<1} \: .
	\label{sec:app_BE:eq:rate_calculation_1}
\end{equation}
Here, $P(\Omega)\big|_{Q<1}$ is given by Eq.~\eqref{sec:app_BE:eq:asymptotics_1}
and we anticipate that the integrand contains the scale $\alpha_g N \ln 2$
that separates the logarithmic divergence at small frequency from the rest.
The first part in Eq.~\eqref{sec:app_BE:eq:rate_calculation_1} yields,
\begin{equation}
	\begin{split}
		\int_{0}^{\alpha_g N \ln 2}\dd\Omega \: P(\Omega)\big|_{Q<1}
		= &
		4\ln 2 \: \alpha_g^2 \pi^2 N \int_{0}^{\alpha_g N \ln 2}\dd\Omega \: \left\{
			\frac{1}{(\alpha_g N\ln 2)^2}\ln\frac{\alpha_g N \ln 2}{|\Omega|}
			+\int_{\alpha_g N\ln2}^1\dd Q\frac{1}{Q(Q^2-\Omega^2)}
		\right\} \\
		= &
		4 \alpha_g \pi^2 (1+\ln 2)
		\: .
	\end{split}
	\label{sec:app_BE:eq:rate_calculation_2}
\end{equation}
The second part in Eq.~\eqref{sec:app_BE:eq:rate_calculation_1} yields
\begin{equation}
	\int_{\alpha_g N \ln 2}^{1}\dd\Omega \: P(\Omega)\big|_{Q<1}
	=
	4\ln 2 \: \alpha_g^2 \pi^2 N \int_{\alpha_g N \ln 2}^{1}\dd\Omega \:
	\frac{\text{arccot}(\alpha_g N\ln2 \Omega)-\arctan(\Omega/\alpha_g N\ln 2)}{2\alpha_g N \ln2 \Omega}
	=
	4 \alpha_g \pi^2 G/2 \: ,
	\label{sec:app_BE:eq:rate_calculation_3}
\end{equation}
where $G=0.916$ is the Catalan constant. Together,
Eqs.~\eqref{sec:app_BE:eq:rate_calculation_2} and \eqref{sec:app_BE:eq:rate_calculation_3}
yield the result~\eqref{sec:cascade_step:eq:Gamma_mu_zero} from the main text.
\subsubsection{The limit $|\mu|\gg T$}
\noindent
In the case $|\mu|\gg T$ we find that the rate $\Gamma$ is determined by small
energy and momentum transfer, $|\Omega|,Q<\alpha_g N|\tilde\mu|/2$.
\begin{equation}
	\Gamma=2T\int_{0}^{\alpha_g N|\tilde\mu|/2}\dd\Omega\:\left(\frac{\alpha_g N|\tilde\mu|}{2}\right)^{-2}
	\int_{|\Omega|}^{\alpha_g N |\tilde\mu|/2}\frac{\dd Q}{Q}
	=8\alpha_g\pi^2|\mu|
\end{equation}

\section{The polarization operator in graphene \label{sec:app_RPA}}
\noindent
We use the dimensionless variables introduced in the preceding sections.
Starting from the definition of the polarization operator in the Keldysh technique\cite{Schuett2011}
\begin{equation}
	\Pi^R=\frac{\ii}{2}\int(\dd\ep)\text{Tr}\left[\hat{G}^R(\ep)\hat{G}^K(\ep+\omega)+\hat{G}^K(\ep)\hat{G}^A(\ep+\omega)\right] \: ,
\end{equation}
we obtain the following expressions for $\Pi^R$ for arbitrary chemical potential and temperature,
\begin{equation}
	\begin{split}
		\text{Im}\Pi^R = & \frac{T Q}{8\pi} \bigg\{\frac{\Theta(1-|\beta|)}{\sqrt{1-\beta^2}}\int_{1}^\infty\dd\xi\sum_{s=\pm1}\sqrt{\xi^2-1}\frac{\sinh(\beta Q)}{\cosh(\beta Q)+\cosh(s\xi Q-{\tilde\mu})} \\
		& - \frac{\Theta(|\beta|-1)}{\sqrt{\beta^2-1}}\int_{-1}^{1}\dd\eta \sqrt{1-\eta^2}\frac{\sinh(\beta Q)}{\cosh(\beta Q)+\cosh(\sgn(\beta)\eta Q+{\tilde\mu})}
		\bigg\} \: ,
	\end{split}
	\label{sec:app_RPA:eq:ImPi}
\end{equation}
\begin{equation}
	\begin{split}
		\text{Re}\Pi^R = & -\frac{T Q}{8\pi^2} P\int_{-1}^{1}\dd\eta\int_{1}^\infty\dd\xi\sum_{s=\pm1}
		\bigg\{ \frac{1}{\beta-s\eta}\sqrt{\frac{\xi^2-1}{1-\eta^2}}\frac{\sinh(s\eta Q)}{\cosh(s\eta Q)+\cosh(s\xi Q-{\tilde\mu})} \\
		& - \frac{1}{\beta-s\xi}\sqrt{\frac{1-\eta^2}{\xi^2-1}}\frac{\sinh(s\xi Q)}{\cosh(\xi Q)+\cosh(s\eta Q+{\tilde\mu})}
		\bigg\} \: .
	\end{split}
	\label{sec:app_RPA:eq:RePi}
\end{equation}
Here $P\int\dots$ denotes the principal value. The asymptotics for $|\tilde\mu|\gg 1$ in all relevant integration regions
are given in Tab.~\ref{sec:app_RPA:tab:PolOp}.
For $|\tilde\mu|\ll1$ they can be found in Ref.~\onlinecite{Schuett2011}.

\begin{table}[h]
\caption{The asymptotics of the polarization operator in graphene
	for $|\tilde\mu|\gg 1$ in the different regimes from
	Eqs.~\eqref{sec:app_RPA:eq:ImPi}~and~\eqref{sec:app_RPA:eq:RePi}.
	Here $I_\eta(z)$ denotes the modified Bessel function of the first kind.
	\label{sec:app_RPA:tab:PolOp}}
\begin{ruledtabular}
\begin{tabular}{c|c|c|c|c|}
	\multirow{2}{*}{} & \multicolumn{2}{c|}{$|\beta|<1$} & \multicolumn{2}{c|}{$|\beta|>1$} \\ \cline{1-5}
	& $Q\ll\tilde\mu$ & $Q\gg\tilde\mu$ & $Q\ll\tilde\mu$ & $Q\gg\tilde\mu$ \\ \hline
	$\text{Re}\Pi^R$
		& $\frac{|\mu|}{2\pi}$
		& $\frac{TQ}{16\sqrt{1-\beta^2}}$
		& $-\frac{T}{8\pi}\: \frac{I_1(Q)}{\beta^2} \: \frac{{\tilde\mu}^2}{Q}$
		& $-\frac{T}{4\pi\beta^2 Q}$ \\ \hline
	$\text{Im}\Pi^R$
		& $\frac{|\mu|}{2\pi} \: \frac{\Omega}{\sqrt{Q^2-\Omega^2}}$
		& $\frac{T}{4\sqrt{2\pi Q}}\:\ee^{-(1-\beta)Q}$, for $(1-\beta)Q\gg 1$
		& $-\frac{T}{16}\frac{Q^2}{\sqrt{\Omega^2-Q^2}}\:
			\frac{\sinh(\Omega)}{\cosh(\Omega)+\cosh(\tilde\mu)}$
		& $-\frac{T}{16}\frac{Q^2 \tanh\Omega}{\sqrt{\Omega^2-Q^2}}$
\end{tabular}
\end{ruledtabular}
\end{table}
\section{Phase space of two particle scattering - the kinetic kernel \label{sec:app_K}}
\noindent
Finally we give the asymptotics of the kinetic kernel
\begin{equation}
	\mathcal{K}(\Omega,Q) = 2T^2 \ee^\Omega \int_{-\infty}^{+\infty} \dd \xi \:
	\frac{ \text{Re}[\sgn(1-|\beta|)(\xi^2-Q^2)]^{1/2} }{4\cosh\frac{\xi-\Omega-\tilde\mu}{2}\cosh\frac{\xi+\Omega-\tilde\mu}{2}} \: ,
	\label{app:kinKernel:eq:kin_Kernel}
\end{equation}
for all integration regions.
\begin{table}[h]
\caption{The asymptotics of the kinetic kernel~\eqref{app:kinKernel:eq:kin_Kernel}
	expressing the phase space for the thermal electrons participating
	at the scattering event.
	\label{sec:app_RPA:tab:kinKernel}}
\begin{ruledtabular}
\begin{tabular}{|c|c|c|c|c|}
	\multirow{2}{*}{$|\tilde\mu|\gg 1$} & \multicolumn{2}{c|}{$\beta<1$} & \multicolumn{2}{c|}{$\beta>1$} \\ \cline{2-5}
	& $Q\ll\tilde\mu$ & $Q\gg\tilde\mu$ & $Q\ll\tilde\mu$ & $Q\gg\tilde\mu$ \\ \hline
	$\mathcal{K}$
		& $4T^2\Omega|\tilde\mu|(1+\coth\Omega)$
		& $2T^2\sqrt{2\pi Q}\ee^{-(1-\beta)Q}$
		& $T^2 \pi Q^2 \ee^{-(1-\sgn(\Omega))|\Omega|}$
		& $T^2 \pi Q^2 \ee^{-(1-\sgn(\Omega))|\Omega|}$
\end{tabular}
\begin{tabular}{|c|c|c|c|c|}
	\multirow{2}{*}{$|\tilde\mu|\ll 1$} & \multicolumn{2}{c|}{$\beta<1$} & \multicolumn{2}{c|}{$\beta>1$} \\ \cline{2-5}
	& $Q\ll 1$ & $Q\gg 1$ & $Q\ll 1$ & $Q\gg 1$ \\ \hline
	$\mathcal{K}$
		& $4T^2\ln 2 \ee^{\beta Q}$
		& $2T^2\sqrt{2\pi Q}\ee^{-(1-\beta)Q}$
		& $T^2 \pi Q^2 \ee^{-(1-\sgn(\Omega))|\Omega|}$
		& $T^2 \pi  Q^2 \ee^{-(1-\sgn(\Omega))|\Omega|}$
\end{tabular}
\end{ruledtabular}
\end{table}
\section{Estimate of the scattering rate from Auger processes \label{sec:app_Auger}}
\noindent
\begin{figure}[h]
	\includegraphics[width=0.75\linewidth]{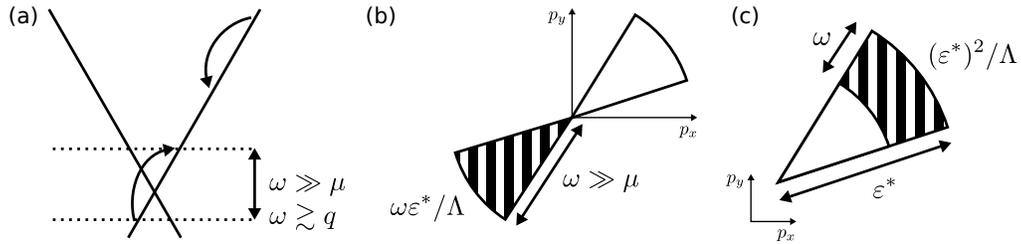}
	\caption{(a) Auger process and (b) phase space for low energy electrons
	in the $p_x$-$p_y$ plane;
	(c) phase space for high energy electron.
	\label{sec:app_Auger:fig:Auger}
	}
\end{figure}
The phase space for Auger processes is controled by the parameter
$\ep^*/\Lambda$, which describes the curvature.
To estimate the contribution to the JSD from Auger processes,
Fig.~\ref{sec:app_Auger:fig:Auger}(a),
we need the phase space for the high energy electron,
which is given by $\sim \omega(\ep^*)^2/\Lambda$ 
[Fig.~\ref{sec:app_Auger:fig:Auger}(c)],
and the phase space for the thermal low energy electrons
$\sim \omega^2\ep^*/\Lambda$ [Fig.~\ref{sec:app_Auger:fig:Auger}(b)].
Their product multiplied by the matrix element of scattering gives
the following estimate for the JSD due to Auger processes,
\begin{equation}
	P_\text{Auger}(\omega)
	=
	\frac{\omega^3(\ep^*)^3}{\Lambda^2}
	\left[\frac{|V(\omega,q)|^2}{|\omega^2-q^2|}\right]_{\omega\gtrsim q}
	\sim \frac{(\ep^*)^3}{\Lambda^2\omega}
	\: .
	\label{sec:app_Auger:eq:P_Auger}
\end{equation}
Comparing $P_\text{Auger}$ with $P$ due to intraband transitions we find that for
$\omega\gtrsim T (T^{1/3}\Lambda^{2/3}/\ep^*)^{2}$, Auger processes dominate.
However, if this threshold lies beyond the particle energy $\ep^*$ we can neglect
them, i.e. for $\ep^*\lesssim \Lambda (T/\Lambda)^{5/9}$.
This applies irrespective of the relation between $T$ and $\mu$, provided $\omega\gg|\mu|,T$.
\end{document}